\documentclass[amsmath,amssymb,showpacs,floatfix]{revtex4}
\usepackage{graphicx}
\usepackage{latexsym}
\begin{document}

\title{Noise enhancing the classical information capacity of a
quantum channel}

\author{Garry Bowen$^{a,}$\footnote{Present address: Centre for
Quantum Computation,
DAMTP, University of Cambridge, Cambridge CB3 0WA, United Kingdom}
and Stefano Mancini$^{b}$}
\affiliation{$^a$Department of Computer Science,
University of Warwick, Coventry CV4 7AL, United Kingdom\\
$^b$INFM \& Physics Department, University of Camerino, I-62032 Camerino,
Italy}

\date{\today}

\begin{abstract}
We present a simple model of quantum communication where
a noisy quantum channel may benefit from
the addition of further noise at the decoding stage.
We demonstrate enhancement of the classical
information capacity of an amplitude damping channel, with
a predetermined detection threshold, by the addition of noise
in the decoding measurement.
\end{abstract}


\maketitle

\section{Introduction}

In common wisdom, randomness and noise are thought to be
detrimental to any physical process, especially at a quantum
level, and hence also for the
ability of the system to encode and process quantum information
\cite{Nie00}. For
classical systems the advent of phenomena such as stochastic
resonance, Brownian ratchets, and Parrondo effects, have shown that
noise may indeed play a helpful role after all \cite{Gam98,Par00}.
This opens up the intriguing question of whether randomness can play
a useful role in quantum systems \cite{Man02},
particularly due to the widespread
current interest in quantum information schemes \cite{Lee02}.

In this paper we consider the quantum communication scenario,
and present a toy model
where the noise plays a positive role.
In particular, given a noisy quantum channel,
we shall show that the addition of noise at the decoding stage
can improve the channel capacity.

\section{The model}

\begin{figure}[h]
\begin{center}
\includegraphics[height=2.0in, width=4.5in ] {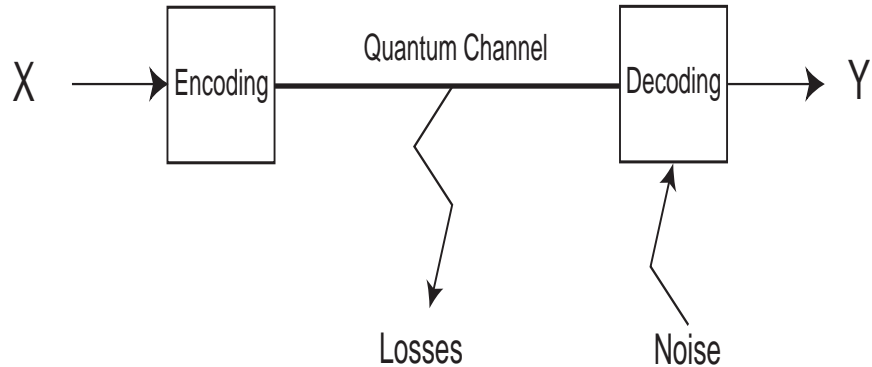}
\end{center}
\caption{Schematic representation of the channel under study.}
\label{fig0}
\end{figure}

Let us consider a simple communication model as depicted in
Fig.\ref{fig0}.
At the sending station, classical information is encoded
into quantum states, which in turn are sent through a noisy quantum
channel.
At the receiving station a decoding procedure is performed with the
addition of noise to the decoding operation.

In detail, we consider an input random binary variable $X$
taking values $0$, $1$ and we encode it into quantum states according
to the
following rule,
\begin{equation}\label{enc}
\left\{\begin{array}{l}
0\longrightarrow
{\hat\rho}_{0}=|0\rangle\langle 0|
\\
1\longrightarrow
{\hat\rho}_{1}=e^{-r^{2}}\sum_{n}\frac{r^{2n}}{n!} |n\rangle\langle n|
\end{array}
\right.\,,
\end{equation}
where ${\hat\rho}_{1}$ is a completely dephased coherent state of
amplitude $r\ne 0$.
We then take the noisy quantum channel to be an amplitude damping
channel
${\hat{\cal E}}$ \cite{Nie00},
the effect of which can be described by the map $r\to r e^{-\tau}$,
with
$\tau$ a parameter determining the amount of amplitude damping.
Finally, the output variable $Y$ is determined by a decoding
procedure involving a quadrature
measurement, through the Probability Operator Measure (POM)
$d{\hat\Pi}(y)=\delta\left(y-{\hat y}\right)dy$
\cite{Hel76,Dar97}.
The binary character of $Y$ comes out from a
threshold detection of the type,
\begin{equation}\label{dec}
\left\{\begin{array}{l}
| y|< \theta\longrightarrow 0
\\
| y|> \theta\longrightarrow 1
\end{array}
\right.\,,
\end{equation}
where $\theta$ represents a fixed threshold.

At this point, additional noise may be introduced at the decoding
stage,
such that we obtain a broadening of the POM \cite{Dar97}, that is,
\begin{equation}\label{pov}
d{\hat\Pi}(y)=\frac{1}{\sqrt{\pi\sigma^{2}}}
\exp\left[-\frac{\left(y-{\hat y}\right)^{2}}{\sigma^{2}}\right]dy\,,
\end{equation}
where $\sigma$ is the width of the Gaussian noise.
For $\sigma\to 0$ the measurement converges to the ideal POM.
The above kind of noise may be used to describe detection with
non-unit
efficiency \cite{Dar97}.

The output conditional probability between $Y$ and the input
amplitude $r$ can be formally written as,
\begin{eqnarray}
p(y|r)&=&{\rm Tr}\left\{{\hat{\cal E}}({\hat\rho})\int
d{\hat\Pi}\right\}
\label{iop}\\
&=&\int dy' \langle y'|{\hat{\cal E}}({\hat\rho})|y'\rangle
\frac{1}{\sqrt{\pi\sigma^{2}}}
\exp\left[-\frac{\left(y-y' \right)^{2}}{\sigma^{2}}\right]\,,
\nonumber
\end{eqnarray}
with,
\begin{equation}\label{yEy}
\langle y'|{\hat{\cal E}}({\hat\rho})|y' \rangle
=\left(\frac{1}{\pi}\right)^{1/2}
\exp\left[-(y')^{2}-r^{2}e^{-2\tau}\right]
\sum_{n}\frac{\left(r^{2}e^{-2\tau}/2\right)^n}{(n!)^{2}}
H_{n}^{2}(y')\,,
\end{equation}
where $H_{n}(.)$ denotes the Hermite polynomial of degree $n$.
Explicitly, one has,
\begin{eqnarray}
p(y|r)&=&\frac{1}{\sqrt{\pi(1+\sigma^{2})}}
\exp\left[
-r^{2}e^{-2\tau}-\frac{y^{2}}{1+\sigma^{2}}\right]
\nonumber\\
&&\times\sum_{n=0}^{\infty}\sum_{k=0}^{n}
\frac{\left(r^{2}e^{-2\tau}/2\right)^n}{\left[(n-k)!\right]^{2}}
\frac{2^k}{k!}\left(\frac{1}{1+\sigma^{2}}\right)^{n-k}
H_{2n-2k}\left(\frac{y}{\sqrt{1+\sigma^{2}}}\right)\,.
\label{pyr}
\end{eqnarray}

By virtue of Eqs.(\ref{enc}) and (\ref{dec}), we obtain the binary
transition probabilities for the input-output states,
\begin{eqnarray}
p_{00}&=&\int_{I}p(y|0)dy\,,
\label{p00def}\\
p_{01}&=&\int_{I}p(y|r)dy\,,
\label{p01def}\\
p_{10}&=&\int_{{\bf R}-I}p(y|0)dy=1-p_{00}\,,
\label{p10def}\\
p_{11}&=&\int_{{\bf R}-I}p(y|r)dy=1-p_{01}\,,
\label{p11def}
\end{eqnarray}
where $I\equiv (-\theta,\theta)$.
Expanding the expressions for the transition probabilities leads to,
\begin{eqnarray}
p_{00}&=&{\rm Erf}\left(\frac{\theta}{\sqrt{1+\sigma^{2}}}\right)\,,
\label{p00}\\
p_{01}&=&\frac{2}{\sqrt{\pi}}
\exp\left[
-r^{2}e^{-2\tau}\right]
\sum_{n=0}^{\infty}\sum_{k=0}^{n}
\frac{\left(r^{2}e^{-2\tau}/2\right)^n}{\left[(n-k)!\right]^{2}}
\frac{2^k}{k!}\left(\frac{1}{1+\sigma^{2}}\right)^{n-k}
\nonumber\\
&&\times\left[
H_{2n-2k-1}(0)-
e^{-\theta^{2}/(1+\sigma^{2})}
H_{2n-2k-1}\left(\frac{\theta}{\sqrt{1+\sigma^{2}}}\right)\right]\,.
\label{p01}
\end{eqnarray}
with the other two transistion probabilities determined according to
Eqs.(\ref{p10def}) and (\ref{p11def}).  The maximum rates of
information transfer may then be determined by maximizing over the
input distribution.

\section{Channel capacity}

Let $p_{0}$ and $p_{1}=1-p_{0}$ the input probabilities
for the values $0$ and $1$, respectively.  The output
entropy of the channel is given by the Shannon entropy of the
variable $Y$ \cite{Nie00}, that is
\begin{equation}\label{HY}
H(Y)={\cal H}\Big(p_{11}p_{1}+(1-p_{00})(1-p_{1})\Big)
+{\cal H}\Big((1-p_{11})p_{1}+p_{00}(1-p_{1})\Big)\,,
\end{equation}
with the function ${\cal H}(z)=-z\log(z)$.
Furthermore, the conditional entropy for $Y$ given $X$ is determined
by \cite{Nie00},
\begin{equation}\label{HYX}
H(Y|X)=(1-p_{1}){\cal H}(p_{00})
+(1-p_{1}){\cal H}(1-p_{00})
+p_{1}{\cal H}(p_{11})+p_{1}{\cal H}(1-p_{11})\,.
\end{equation}
The channel capacity may then be obtained by maximizing the mutual
information
$H(Y)-H(Y|X)$, over all possible input probabilities \cite{Nie00},
explicitly,
\begin{equation}\label{Cdef}
C=\max_{\{p_{1}\}}\left[H(Y)-H(Y|X)\right]\,.
\end{equation}
The input probability maximizing Eq.(\ref{Cdef}) may be shown to be,
\begin{equation}\label{P1st}
p_{1}^{*}=\frac{p_{00}-\wp}{p_{00}+p_{11}-1}\,,
\end{equation}
with,
\begin{equation}\label{wpinv}
\wp^{-1}=1+\exp\left[
\ln(2) \;\frac{{\cal H}(p_{00})+{\cal H}(1-p_{00})
-{\cal H}(p_{11})-{\cal H}(1-p_{11})}
{p_{00}+p_{11}-1}
\right]\,.
\end{equation}

\begin{figure}[htbp]
\begin{center}
\includegraphics[height=2.5in, width=3.5in ] {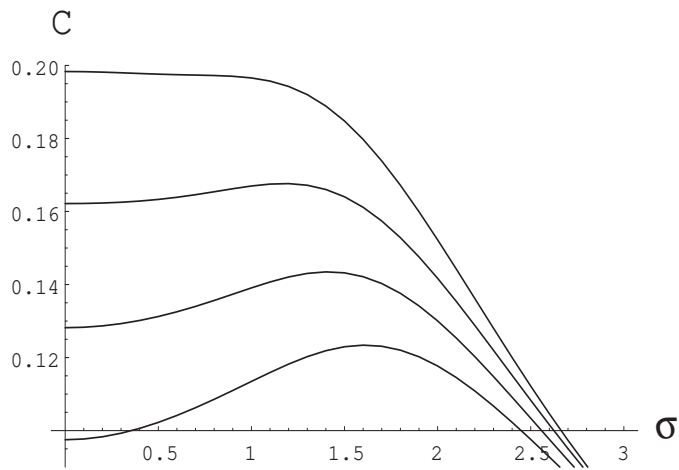}
\end{center}
\caption{Channel capacity as function of the noise strength $\sigma$.
Curves from top to bottom are for threshold values $\theta=3.6$,
$3.8$, $4.0$ and $4.2$.
The value of amplitude and damping parameters are $r=5$ and
$\tau=0.5$.
}
\label{fig1}
\end{figure}

In Fig.\ref{fig1} we graph the capacity $C$ versus the noise strength
$\sigma$ for different values of the threshold $\theta$, when the
encoding
amplitude and channel damping are fixed to $r=5$ and $\tau=0.5$
respectively.
One can see that when the threshold $\theta$ becomes higher than
the damped amplitude $re^{-\tau}$, the addition of the noise
improves the distinguishability between the two outputs
in the decoding stage, thus
increasing the capacity. Clearly, there is an optimal strength value
for the noise to be added, beyond which
the capacity goes to zero.

\section{Conclusion}

In conclusion, we have shown
the noise enhancement of the classical information
capacity of a quantum communication channel.
Analogous results have previously been found within the classical
context \cite{Cha97}.
The effect can be ascribed to the ability of the noise to
induce transitions on the threshold \cite{Gam98},
and in our case it results from competition between
two differing types of noise \cite{Man01}, the noise
causing losses in the channel, and the noise broadening the POM.

The noise benefit could be optimized by a suitable tailoring
of the POM and of the involved parameters,
but this requires large numerical resources.
Nonetheless, the present simple model
can serve as a proof of principle and
as a useful tool for further developments
in more realistic schemes.
In particular it could be relevant for systems that are
constrained to receive signals through a fixed threshold \cite{Hel76}.

Our approach paves the way for a thorough study of noise effects in
nonlinear quantum channels.
Furthermore, the addition of stochastic noise on a channel may
be helpful for quantum binary decision,
which find applications in a variety of quantum measurement problems
\cite{Hel76}.

On the other hand, quantum correlations (entanglement)
have been exploited to improve both channel capacity \cite{Mac02}
and quantum binary decision \cite{Dar02}.
It would be interesting to compare the two approaches
and to eventually combine them.

\end{document}